\documentstyle[12pt,aaspp]{article}
  at 12.0 true pt 

\def\etal{{et al}}

\def\phoenix{{\tt PHOENIX}}

\def\g{\gamma}
\def\b{\beta}
\def\m{\mu}

\def\n{\nu}

\def\pder#1#2{{\partial #1 \over \partial #2}}
\def\div#1#2{{#1\over #2}}
\def\rout{\ifmmode{r_{\rm out}}\else\hbox{$r_{\rm out}$}\fi}
\def\tmax{\ifmmode{\tau_{\rm max}}\else\hbox{$\tau_{\rm max}$}\fi}
\def\tstd{\ifmmode{\tau_{\rm std}}\else\hbox{$\tau_{\rm std}$}\fi}
\def\vmax{\ifmmode{v_{\rm max}}\else\hbox{$v_{\rm max}$}\fi}
\def\muE{\ifmmode{\mu_{\rm E}}\else\hbox{$\mu_{\rm E}$}\fi} 
\def\pE{\ifmmode{p_{\rm E}}\else\hbox{$p_{\rm E}$}\fi} 
\def\bmax{\ifmmode{\b_{\rm max}}\else\hbox{$\b_{\rm max}$}\fi}
\def\kms{\hbox{$\,$km$\,$s$^{-1}$}}

\def\ang{\hbox{\AA}}

\def\Lsun{\hbox{$\,$L$_\odot$} }
\def\Teff{\hbox{$\,T_{\rm eff}$} }

\def\rout{\hbox{$r_{\rm out}$} }
\def\pgas{\hbox{$P_{\rm gas}$} }
\def\chistd{\ifmmode{\chi_{\rm std}}\else\hbox{$\chi_{\rm std}$}\fi}

\def\doublet#1{\hbox{$ ^2$#1}}

\def\lstar{\ifmmode{\Lambda^*}\else\hbox{$\Lambda^*$}\fi} 
\def\Rop{\ifmmode{[R_{ij}]}\else\hbox{$[R_{ij}]$}\fi}

\def\Rji{\ifmmode{[R_{ji}]}\else\hbox{$[R_{ji}]$}\fi}
\def\Rstar{\ifmmode{[R_{ij}^*]}\else\hbox{$[R_{ij}^*]$}\fi}

\def\Rjistar{\ifmmode{[R_{ji}^*]}\else\hbox{$[R_{ji}^*]$}\fi}
\def\DRji{\ifmmode{[\Delta R_{ji}]}\else\hbox{$[\Delta R_{ji}]$}\fi}
\def\DRij{\ifmmode{[\Delta R_{ij}]}\else\hbox{$[\Delta R_{ij}]$}\fi}

\def\ns{\ifmmode{N_{\rm s}}          
        \else\hbox{$N_{\rm s}$}\fi}

\def\mat#1{{\bf #1}}     
\def\vek#1{{#1}}         

\newcount\eqcount
\def
  \nummer{
    \global\advance\eqcount by 1
    (\the\eqcount)
  }

\def
  \numadv{
    \global\advance\eqcount by 1
  }

\def
   \numout#1{
     (\the\eqcount #1)
  }

\def\ivek#1#2{\ifmmode{\vek{I}^{#1}_{#2}}
        \else\hbox{$\vek{I}^{#1}_{#2}$}\fi}

\def\tmat#1#2{\ifmmode{\mat{t}^{#1}_{#2}}
        \else\hbox{$\mat{t}^{#1}_{#2}$}\fi}
\def\rmat#1#2{\ifmmode{\mat{r}^{#1}_{#2}}
        \else\hbox{$\mat{r}^{#1}_{#2}$}\fi}
\def\bvek#1#2{\ifmmode{\beta^{#1}_{#2}}
        \else\hbox{$\beta^{#1}_{#2}$}\fi}

\def\lp{\ifmmode{\lambda^+_\tau}           
        \else\hbox{$\lambda^+_\tau$}\fi}
\def\lm{\ifmmode\lambda^-_\tau             
        \else\hbox{$\lambda^-_\tau$}\fi}

\def\la{\mathrel{\hbox{\rlap{\hbox{\lower4pt\hbox{$\sim$}}}\hbox{$<$}}}}
\def\ga{\mathrel{\hbox{\rlap{\hbox{\lower4pt\hbox{$\sim$}}}\hbox{$>$}}}}

\def\bea{\begin{eqnarray}}
\def\eea{\end{eqnarray}}
\def\be{\begin{equation}}
\def\ee{\end{equation}}

\def\etal{et al.}

\def\fep{\hbox{Fe II} }

\begin{document}
\bibliographystyle{apj}

\title{The Effects of Fe~II NLTE on Nova Atmospheres and Spectra}

\author{Peter H. Hauschildt}
\affil{Dept.\ of Physics and Astronomy, Arizona State University, Box 871504,\\
Tempe, AZ 85287-1504\\
E-Mail: \tt yeti@sara.la.asu.edu}
\author{E. Baron}
\affil{Dept. of Physics and Astronomy, University of Oklahoma,\\
440 W. Brooks, Rm 131, Norman, OK 73019-0225\\
E-Mail: \tt baron@phyast.nhn.uoknor.edu}
%
%
\author{Sumner Starrfield}
\affil{Dept.\ of Physics and Astronomy, Arizona State University, Box 871504,\\
Tempe, AZ 85287-1504\\
E-Mail: \tt starrfie@hydro.la.asu.edu}
\and
\author{France Allard}
\affil{Dept.\ of Physics, Wichita State University,\\
Wichita, KS 67260-0032\\
E-Mail: \tt allard@mugem.phy.twsu.edu}

\baselineskip=24pt

\begin{abstract}

The atmospheres of novae at early times in their outbursts are very extended, 
expanding shells with low densities. Models of these atmospheres show that 
NLTE effects are very important and must be included in realistic 
calculations. We have, therefore, been improving our atmospheric studies by 
increasing the number of ions treated in NLTE. One of the most important ions 
is Fe~II which has a complex structure and numerous lines in the observable 
spectrum. In this paper we investigate NLTE effects for Fe~II for a wide 
variety of parameters. We use a detailed Fe~II model atom with 617 level and 
13675 primary lines, treated using a rate-operator formalism. We show that the 
radiative transfer equation in nova atmospheres {\em must} be treated with 
sophisticated numerical methods and that simple approximations, such as the 
Sobolev method, {\em cannot} be used because of the large number of 
overlapping lines in the co-moving frame. 

 Our results show that the formation of the Fe~II lines is strongly affected 
by NLTE effects. For low effective temperatures, $\Teff < 20,000\,$K, the 
optical Fe~II lines are most influenced by NLTE effects, while for higher 
$\Teff$ the UV lines of Fe~II are very strongly affected by NLTE. The 
departure coefficients are such that Fe~II tends to be overionized in NLTE 
when compared to LTE. Therefore, Fe~II-NLTE must be included with 
sophisticated radiative transfer in nova atmosphere models in order to 
reliably analyze observed nova spectra. Finally, we show that the number of 
wavelength points required for the Fe~II NLTE model atmosphere calculations 
can be reduced from 90,000 to about 30,000  without changing the results if we 
choose a sufficiently dense UV wavelength grid. 

\end{abstract}

\section{Introduction}


The detailed modeling of early nova atmospheres and spectra has made
significant progress over the last few years (cf.\ Hauschildt \etal\
1992, 1995\nocite{novapap,novaphys}). Although the fits of synthetic
spectra computed from model atmospheres to observed nova
spectra is in general satisfactory (Hauschildt \etal,
1994ab)\nocite{cygpap,cas93pap}, there are still a number of
discrepancies between synthetic and observed  nova spectra. In
particular in the ``pre-nebular'' phase of the nova evolution, when the
effective temperature of the nova atmosphere is about $20000$ to
$30000\,$K, the fits to the \fep lines in the UV spectral region (from
$1000$ to $3500\ang$) become much worse than fits obtained in the
earlier ``optically thick wind'' ($\Teff \approx 10000$--$20000\,$K)
phase. In general, we have found that the lines of \fep are too strong
in the synthetic spectra for the pre-nebular phase when compared to
the observations. However, the same \fep lines and ``clusters'' of lines
(i.e., groups of lines very close in wavelength) are
reproduced well in models for the optically thick wind phase. In
this paper we show that these problems can be resolved by including a
detailed NLTE treatment of the important \fep ion both in the nova model
atmosphere calculations and the synthetic spectra.


 The basic physical background of the formation of early nova spectra
has been discussed  in detail by Hauschildt \etal\ (1995, hereafter,
Paper~I)\nocite{novaphys}. One of their most important results is that
the formation of early nova spectra is extremely complex and that it
cannot be described or modeled accurately using simple approximations.
Nova spectra are formed in an environment which has an enormous density of
lines, i.e., many
thousands of spectral lines overlap in the co-moving frame and,
in addition, continuum absorption and scattering processes are
important. {\em Therefore, the simple Sobolev approximation cannot
be used to replace an accurate solution of the radiative
transfer equation} (Baron \etal\ 1995)\nocite{advecpap}

 In our previous studies we have found that NLTE effects are also very
important for the formation of early nova spectra. This means
that
multi-level NLTE rate equations must be solved for a number of species
self-consistently with the radiative transfer and energy equations.
The effects of line blanketing and the differential velocity
field of the expanding nova shell must also be included for an accurate
treatment of the emergent radiation. So far, we have included only a
limited number of the important species in full NLTE, for the remaining
species and spectral lines we had to use LTE occupation numbers. An
approximate treatment of line scattering was made in order to keep the
nova atmosphere problem technically feasible. We have found that all
species treated in NLTE show significant deviations from LTE in both their
ionization balance and their line formation processes. Therefore, it is
important to investigate the NLTE effects of one of the most important
ions (in terms of line opacity) found in nova atmospheres, the \fep ion.

 Hauschildt \& Baron (1995, hereafter HB95)\nocite{fe2pap} have used the
numerical method developed by Hauschildt (1993, hereafter,
H93)\nocite{casspap} for NLTE calculations with a very detailed model
atom of \fep\unskip. In this paper we will apply this method to nova
model atmospheres and discuss the results and implications for the
analysis of early nova spectra.


 In the following section we will describe briefly the \fep model atom
and the basic features of nova model atmospheres (details can be
found in Paper I and HB95). We will then discuss the results we have
obtained for representative nova model atmospheres, in particular the
effects of \fep NLTE on the ionization balance and the formation of the
\fep lines. We conclude with a summary and discussion.

\section{Methods and Models}


\subsection{Radiative transfer}

In the Lagrangian frame (``comoving frame''),
the special relativistic equation of radiative transfer is given by
(\cite{FRH})
\begin{eqnarray}
%
  \g (\mu + \b) \pder{I}{r}
%
   &  + \pder{}{\mu}\left\{ \g (1-\mu^2)
      \left[ {(1+\b\mu) \over r}
                - \g^2(\mu+\b) \pder{\b}{r} \right] I \right\} \nonumber \\
  &
    - \pder{}{\n} \left\{ \g
       \left[ \div{\b(1-\mu^2)}{r} + \g^2\mu(\mu+\b)\pder{\b}{r} \right]
         \n I \right\} \nonumber \\
  &
    + \g\left\{
       \div{2\mu+\b(3-\mu^2)}{r} + \g^2(1+\mu^2+2\b\mu)\pder{\b}{r}\right\}
      I \nonumber \\
  & \qquad =
      \tilde\eta - \chi I.\label{RTE}
\end{eqnarray}
Here, $r$ is the radius, $\m$ the cosine of the angle between a
ray and the direction normal to the surface, $\n$ the frequency,
$I=I(r,\m,\n)$ denotes the specific intensity
at radius $r$ and frequency $\n$ in the direction $\arccos\m$ in the
Lagrangian frame.
The matter velocity $v(r)$ is measured in units of the speed of light
$c$: $\b(r)=v(r)/c$ and $\g$ is given by $\g(r)=1/\sqrt{1-\b^2}$.
The sources of radiation present in the
matter are described by the emission coefficient
$\tilde\eta=\tilde\eta(r,\n)$,
and $\chi=\chi(r,\n)$ is the extinction coefficient.


We solve Eq.~(\ref{RTE}) using an operator splitting method with a non-local
band-matrix approximate $\Lambda$-operator (\cite{s3pap,aliperf}).
As shown by Bath and Shaviv (1976\nocite{bath76}), we can safely neglect
the {\em explicit} time dependencies and {\em partial} time derivatives
$\partial/\partial t$ in the radiative  transfer because the radiation-
flow timescale is much smaller than the timescale of the evolution of
the nova outburst (even in the fireball stage). However, the advection
and aberration terms must be retained in order to obtain a physically
consistent solution (\cite{FRH}). This approach is also consistent
with the equation of radiation hydrodynamics in the time independent
limit (\cite{FRH,advecpap}).  We neither neglect the Lagrangian time
derivative $D/Dt = \partial/\partial t + v \partial/\partial r$,  nor
assign an ad hoc value to $D/Dt$ as done by Eastman and Pinto
(1993)\nocite{pee}. The latter two assumptions lead to physical
inconsistencies with the equations of radiation hydrodynamics (e.g.,
they do not lead to the correct equations for a steady-state stellar
wind). Our approach is physically self-consistent because it includes
the important advection and aberration terms in both the non-grey
radiative transfer and the radiative energy equations. The {\em only}
term that we neglect is the {\em explicit} time dependence, which is a
very good approximation in nova and supernova atmospheres (Baron \etal, 
1995)\nocite{advecpap}.

Although our method is much more complicated than ad hoc assumptions for 
$D/Dt$ (because of the additional terms in the equations that must be handled 
that break the symmetry of the characteristics of the radiative transfer 
equation), its results are much more reliable than those of simpler methods. 
In addition, the solution of the correct set of radiative transfer and energy 
equations in the co-moving frame is no more time consuming than the solution 
of the corresponding static problem. This is because of our use of a non-local 
approximate $\Lambda$-operator. 

\subsection{Treatment of spectral lines}

The physical conditions prevailing in nova atmospheres are such that a large
number of spectral lines are present in the line forming regions. Therefore,
the simple Sobolev approximation {\em cannot} be used because many lines of
comparable strength overlap (see Paper I). This means that the basic
assumptions required to derive the Sobolev approximation are not valid. We
demonstrate this in Fig.~\ref{linefig}a in which we plot the number of lines
that are stronger than the local b-f continuum and that lie within a $\pm 2$
Doppler-width wavelength interval around each wavelength point in the co-
moving frame. This graph is for a nova atmosphere with an effective
temperature of $20,000\,$K, solar abundances and a micro-turbulent or
statistical velocity of $\xi=2\kms$. In the UV the number of overlapping
strong lines at each wavelength point is typically larger than 10, in some
regions as many as 60 strong lines lie within 2 Doppler-widths. Even in some
regions of the optical spectral range, the number of overlapping lines can be
as high as 10 or more. In Fig.~\ref{linefig}b we show that the situation
becomes much worse for weaker lines (lines that are stronger than $10^{-4}$ of
the local b-f continuum must be included in nova atmosphere models, cf.\
Paper~I). Now more than 1000 lines (all of comparable strength) in the
UV and  around 100 lines in the optical can overlap at each wavelength
point. This shows that the Sobolev approximation cannot be used in modeling
nova atmospheres. Therefore, we solve the full radiative transfer equation for
all lines.

\section{The Fe~II model atom}



We take the data needed to construct our model atom from the compilation of 
R.~L.~Kurucz (1994)\nocite{cdrom22}, whose long term project to provide 
accurate atomic data for modeling stellar atmospheres is an invaluable service 
to the scientific community. For our current model atom, we have kept terms up 
to the \doublet{H} term, which corresponds to the first 29 terms of \fep. 
Within these terms we include all observed levels that have observed b-b 
transitions with $\log{gf} > -3.0$ as NLTE levels (where $g$ is the 
statistical weight of the lower level and $f$ is the oscillator strength of 
the transition). This leads to a model atom with 617 levels and 13675 
``primary'' transitions which we treat in detailed NLTE. We solve the complete 
b-f \& b-b radiative transfer and rate equations for all these levels and 
include all radiative rates of the primary lines. In addition, we treat the 
opacity and emissivity for the remaining nearly 1.2 million ``secondary'' b-b 
transitions in NLTE, if one level of a secondary transition is included in our 
detailed model. We give a detailed description of the \fep line treatment 
below. 

\subsection{Photo-ionization Rates}

Detailed photo-ionization rates for \fep have yet to be published, although 
this is one of the goals of the iron project. Thus, we have taken the results 
of the Hartree Slater central field calculations of Reilman \& Manson 
(1979)\nocite{reilman79} to scale the ground state photo-ionization rate and 
then have used a hydrogenic approximation for the energy variation of the 
cross section. Although these rates are only very rough approximations, they 
are useful for initial calculations. For the conditions of the test cases we 
consider in this paper, the exact values of the b-f cross sections are not 
important for the opacities (which are dominated by known b-b transitions of 
\fep and other species), but they do have an influence on the actual b-f 
rates.  This is, of course, unimportant for the computational method which we 
use and the b-f cross sections can be changed easily when better data become 
available. 

\subsection{Collisional Rates}

For collisional rates we have had to make rather rough approximations using 
semi-empirical formulae. We have calculated bound-free collisional rates 
using the semi-empirical formula of Drawin (1961)\nocite{drawin61}. The bound-
bound collisional rates are approximated by the semi-empirical formula of 
Allen (1973)\nocite{allen_aq} and, for permitted transitions, we use 
Von~Regemorter's formula (\cite{vr62}). While the collisional rates are 
important in stellar atmospheres with high electron densities, they are nearly 
negligible when compared to the radiative rates in the low density envelopes 
of novae. If better collisional cross sections become available, they can be 
incorporated very easily. 

\subsection{Calculational Method}

The large number of transitions of the \fep ion that have to be included in 
realistic models of the \fep NLTE line formation require an efficient method 
for the numerical solution of the multi-level NLTE radiative transfer problem. 
As already mentioned, the \fep model atom  described here includes more than 
13000 individual NLTE lines plus a large number of weak background 
transitions. Classical techniques, such as the complete linearization or the 
Equivalent Two Level Atom methods, are computationally prohibitive. In 
addition, we are also interested in modeling moving media, therefore, 
approaches such as Anderson's multi-group scheme or extensions of the opacity 
distribution function method (\cite{HubLan95}) cannot be applied. Again, 
simple approximations such as the Sobolev method, are very inaccurate in 
problems in which lines overlap strongly and make a significant continuum 
contribution (important for weak lines), as is the case for nova (and SN) 
atmospheres (cf.\ Paper I). 

 We use, therefore, the multi-level operator splitting method described by
Hauschildt (1993, hereafter H93). This method solves the non-grey,
spherically symmetric, special relativistic equation of radiative
transfer in the co-moving (Lagrangian) frame using the operator
splitting method described in Hauschildt\nocite{s3pap} (1992, hereafter H92). It
has the advantages that (a) it is numerically highly efficient and
accurate, (b) it treats the effects of overlapping lines and continua
(including background opacities by lines and continua not treated
explicitly in NLTE) self-consistently, (c) gives very stable and smooth
convergence even in the extreme case of novae, and (d) it is not
restricted to a certain application but can be applied to a wide variety
of astrophysical problems. Details of the method are described in H92
and H93, so we give here only a summary of the purely technical
improvements necessary to make the NLTE treatment of very large model
atoms more efficient.

\subsection{Full NLTE Treatment: Primary (strong) Lines}

Even with highly effective numerical techniques, the treatment of
possibly more than one million NLTE lines poses a significant
computational problem, in particular in terms of memory usage. In
addition, most lines are very weak and do not contribute
significantly to the radiative rates. However, together, they can
influence the radiation field from overlapping stronger transitions and
should be included as background opacity. Therefore, we separate the
stronger ``primary'' lines from the weaker ``secondary'' lines by
defining a threshold in $\log(gf)$, which can be arbitrarily changed.
Lines with $gf$-values larger than the threshold are treated in detail,
i.e., they are fully included as individual transitions in the radiative
transfer (assuming complete redistribution) and rate equations. In
addition, we include special wavelength points within the profile of the
strong primary lines.

 The secondary transitions are included as background NLTE opacity sources but 
are not explicitly included in the rate equations. Their cumulative effect on 
the rates is included, since the secondary lines are treated by line-by-line 
opacity sampling in the solution of the radiative transfer equation. Note that 
the distinction between primary and secondary transitions is only a matter of 
convenience and technical feasibility. It is {\em not} a restriction of our 
method or the computer code but can be easily changed by 
altering the appropriate input files.  As more powerful computers become 
available, all transitions can be handled as primary lines by simply changing 
the input files accordingly. 

 In a typical calculation, we use a threshold of $\log(gf) = -3$ so that lines 
with $gf$-values larger than this value are treated as primary lines and all 
other lines are treated as secondary lines. We do not pose additional 
thresholds such as the energy or the statistical weight of the lower level of 
a line. However, we include in the selection process only observed lines 
between known levels in order to include only lines with well known $gf$-
values. All predicted lines of Kurucz are included as secondary lines (see 
below). 

 Using this procedure to select our model atom, we obtain 13675 primary NLTE 
lines between the 617 levels included in NLTE. For \fep\ lines with $\lambda > 
3500\ang$, we use 5 to 11 wavelength points within their profiles. In 
extensive test calculations, we have found that \fep\ lines with $\lambda < 
3500\ang$ do not require these additional wavelength points in their profiles 
due to the fine wavelength grid used in the model calculations at these 
wavelengths (HB95). This procedure typically leads to about 30,000 wavelength 
points for the model iteration and the synthetic spectrum calculations. We 
show later in this paper that this wavelength grid is sufficiently dense for 
the calculations by comparing these results with those obtained using about 
90,000 wavelength points. This is mostly due to the properties of the \fep\ 
ion, in particular that its lines are concentrated mostly in the wavelength 
range below $3500\ang$. Other model atoms, e.g., Ti~I require additional 
wavelength points for practically every line (Hauschildt \etal, in 
preparation). For all primary lines the radiative rates and the ``approximate 
rate operators'' (see H93) are computed and included in the iteration process. 

\subsection{Approximate NLTE Treatment: Secondary Lines and LTE Levels}

The vast majority of the 1.2 million \fep lines in the database of
Kurucz are either very weak lines or are predicted lines (sometimes
between predicted or auto-ionizing levels). Although these weak lines
may be a non-negligible contribution to the overall opacity and the shape of the
resulting ``pseudo-continuum,'' they are not very important for the rate
equations. The transitions between the bound states are dominated by the
fewer primary transitions, which we include individually in the
radiative transfer and rate equation solution.

 However, the ``haze'' of weak secondary lines is included as background
opacity (and hence indirectly in the rate equations). This is also true
for the numerous lines of species considered in LTE. Neglecting the
line-blanketing may lead to wrong results for the NLTE departure
coefficients, see, e.g., Hauschildt \& Ensman (1994)\nocite{snlisa} for a
description of this effect in supernova model atmospheres.

 Therefore, we include the opacity of the secondary lines, defined as all
available \fep lines that are not treated as primary lines, as background opacity.
We distinguish between:
\begin{enumerate}
\item lines for which the lower level is an NLTE
level but the upper level is an LTE level,
\item lines for which the upper
level is an NLTE level but the lower level is an LTE level, and,
\item lines
for which both levels are LTE levels.
\end{enumerate}
Here, an 'NLTE level' is
a level that is explicitly treated in the NLTE rate equations whereas an 'LTE
level' is not considered explicitly in the rate equations. In the first two
cases, we set the departure coefficient for the LTE level equal to the
departure coefficient of the ground state,
whereas in the third case we use the same approach as for the
lines of LTE species (Baron \etal\ 1994, Hauschildt \etal\ 1994ab)
\nocite{sn93jpapbig,cygpap,cas93pap} except
that we use the ground state departure
coefficient to include the effects of over or under-ionization.
This approximate treatment of the secondary lines does not significantly
influence the emergent spectra, as the secondary lines are by definition only
relatively weak lines. We have verified in test
calculations (HB95) that the influence of the secondary lines is indeed
negligible for all models presented here. However, they are included, for
completeness, in the models discussed in this paper.

\subsection{The nova atmosphere model}

 The details of our model assumptions and parameterization of nova atmospheres 
are discussed in Paper~I, therefore, we give here only a short summary. The 
radial dependence of the density in the nova atmosphere is assumed to follow a 
power law $\rho\propto r^{-N}$ with a power law index $N\approx 3$. Such a 
relatively flat (compared to stars and supernovae) density gradient is valid 
for novae in the optically thick wind phase. In addition, the atmosphere is 
assumed to be spherically symmetric, which is a necessary assumption to make 
the problem tractable. We compute the temperature structure using radiative 
equilibrium in the co-moving frame. All models reported here are self-
consistent solutions of the radiative transfer, energy and rate equations and 
iterated to full convergence. The most important model parameters are the 
density exponent $N$, the effective temperature $\Teff$, the maximum expansion 
velocity $\vmax$, and the elemental abundances. The velocity field inside the 
atmosphere is computed with the condition $\dot M(r)=\rm const$. While the 
luminosity, $L$, of the nova is a formal parameter of our models,  it 
does not significantly affect the synthetic spectra (see Paper I). 

\section{Results}


Using the methods described in the previous sections, we have re-computed
our nova model atmosphere grid for $L=50,000\Lsun$,
$\vmax=2000\kms$, $N=3$, solar abundances and a set of effective
temperatures from $10,000$ to $60,000\,$K. These models include not only
NLTE for Fe~II, but also for H~I, He~I, O~I, Mg~II, Ne~I, and Ca~II.
The models were computed with our generalized model atmosphere code
\phoenix, Version 5.7. For a detailed description of the code and
the input physics see Paper~I. In the following discussion, we will only
consider a subset of 4 effective temperatures, namely $10,000$,
$15,000$, $20,000$, and $25,000\,$K. These effective temperatures are
representative for the early evolution of a nova and these models are
sufficient to understand the basic NLTE effects of Fe~II in novae.

\subsection{NLTE effects on the Fe~II ionization}


In Fig.~\ref{bground} we display the ground state departure
coefficients, $b_1$, for these four nova models. We have marked the $b_1$ for
Fe~II with connected plus-signs for clarity. In all models, the
departures from LTE are significant. This is in particular true for
Fe~II, which shows the largest deviations from LTE in the ground state
of all the species shown. However, it is interesting to note that Ca~II
exhibits a behavior very similar to Fe~II, at least for
$\Teff<20,000\,$K.

For most regions of the nova atmospheres, the Fe~II $b_1$ are smaller
than unity. This indicates an overionization of Fe~II relative to
LTE. Other species do not show this behavior, e.g., the ground state
departure coefficients for Ne~I tend to be larger than one for most
models and drop below unity only in the outermost regions of the nova
atmosphere. 
This is likely due to both the low
ionization energy of Fe~II, as well as the extreme complexity of
the Fe~II ion.
These regions mark the transition from the nova
atmosphere to the nebula that is formed by the ejecta. In Fig.~\ref{fe-ion}
we display the ionization structure of iron throughout the
atmosphere for the model with $\Teff=20,000\,$K. In panel (a) we show
the NLTE-Fe~II results whereas panel (b) shows the results for Fe~II
treated in LTE (but otherwise the same model structure). First note the
large number of iron ionization stages that are present in the
atmospheres (in a number of interleaved ``Str\"omgren Spheres''). This
is a well-known feature of nova atmospheres which is caused by the large
temperature gradients inside the expanding shell. We find similar
effects for all elements that are included in the model calculations.

The effects of NLTE on the Fe~II ionization balance become clear by comparing 
panels (a) and (b) in Fig.~\ref{fe-ion}. In the LTE case, Fe~III (the dominant 
ionization stage of iron in the outer atmosphere), recombines to Fe~II outside 
of $\tstd\approx 10^{-4}$. This causes strong Fe~II lines to appear in 
synthetic spectra with LTE-Fe~II, see below. In the NLTE-Fe~II case, however, 
the strong UV radiation field prevents Fe~III from recombining into Fe~II 
throughout the outer atmosphere. This is a very important change and it  
results in significantly weaker Fe~II lines than found in the Fe~II LTE 
models. We will discuss the observational implications of these results later 
in more detail. 

 The ground state departure coefficients of all species show a complicated
behavior for $0.01 \le \tstd \le 1$ and this is very obvious in the
model for $\Teff=20,000\,$K. This is caused by the presence of a number of
ionization zones, He~I to He~II, He~II to He~III, and O~III to O~IV, in this
region of the atmosphere (cf.\ Fig.~\ref{ion-struc}). This requires the
departure coefficients to change very rapidly, as they sensitively depend on
the degree of ionization of the species and the electron density.
Therefore, we show in Fig.~\ref{nefrac} the ratio $p({\rm Fe~II})/p({\rm
Fe~III})$ (curve marked with plus-symbols, left hand axis) and the ratio
$p_{\rm elec}/\pgas$ (dotted curve, right hand axis) on the same graph. The
sudden changes in the electron pressure (mainly triggered by the two helium
ionization zones), in combination with the rapid changes in the degree of iron
ionization, cause the behavior of the departure coefficients.

 In Fig.~\ref{allbi} we show the departure coefficients for all \fep\ levels
for the 4 models in an overview graph. This figure
demonstrates that the NLTE effects for \fep\ become more important with higher
effective temperatures. For each $\Teff$, the departure coefficients are
closer to unity (their LTE value) for the higher lying levels, which are
more strongly coupled to the continuum than the lower lying levels.
The rather large variation of the $b_i$ within the \fep\ model atom indicates
that the \fep\ lines will show significant NLTE effects themselves, in
addition to the effects introduced by the changes in the ionization structure.

\subsection{NLTE effects on the formation of Fe~II lines}


 The NLTE effects not only change the ionization balance of iron but also 
change the level populations of Fe~II. We demonstrate this effect for a number 
of important \fep\ multiplets. In Fig.~\ref{uv1} we show the ratio of the line 
source functions $S_L$ to the local Planck function for the UV1 multiplet of 
\fep. These are UV lines (around $2600\ang$) from the ground $\rm a ^6D$ term 
to $\rm z ^6D^o$. In the model with $\Teff=15,000\,$K (Fig.~\ref{uv1}a) the 
UV1 multiplet is nearly in LTE in the outer atmosphere. Only between 
$\tstd=10^{-2}$ and $1$ do the line source functions differ, by about 20\%, 
from the Planck function. This is probably caused by the ionization zone from 
\fep\ to Fe~III which occurs at these optical depths. The situation is very 
different for $\Teff=25,000\,$K (Fig.~\ref{uv1}b). Here, the line source 
functions are smaller than the Planck function by up to a factor of 10. In 
both models, the (collisional) coupling between the levels of each term is 
strong enough to nearly establish the same line source function to Planck 
function, $S_L/B$, ratio for all lines within the multiplet, although the 
electron densities are relatively low. 

 The subordinate \fep-multiplet 42 ($\rm a ^6S$ to $\rm z ^6P^o$) shows
a somewhat different behavior (cf.\ Fig.~\ref{42}). The line source
functions for the 3 primary NLTE lines are smaller than unity throughout
the line forming region for both the $\Teff=15,000$ (Fig.~\ref{42}a) and the
$\Teff=25,000\,$K (Fig.~\ref{42}b) models. The NLTE effects are larger for the
hotter model, as seen earlier for the UV1 multiplet. This shows that the NLTE
effects for \fep\ are not just over-ionization but that the ``internal''
NLTE effects are also significant. We emphasize this point in
Fig.~\ref{from} by showing the $S_L/B$ for transitions from
the ground term $\rm a ^6D$ (panel a), the metastable $\rm a ^4F$ term
(panel b), and the higher $\rm z ^4P^o$ term (panel c) for both the
$15,000$ and the $25,000\,$K models. The general trend is that the NLTE
effects become more important for higher effective temperatures and they
become less important for higher lying terms, as expected.

\subsection{The effects of Fe~II NLTE on nova spectra}


The results that we have described in the preceding sections indicate that
the treatment of Fe~II in NLTE must have significant effects on the spectra
emitted by nova atmospheres. In the following discussion, we distinguish
between cooler ($\Teff <20,000\,$K) and hotter nova models.

\subsubsection{Nova atmospheres with $\Teff <20,000\,$K}

 For nova atmospheres with relatively low effective temperatures, we find that 
the effects of NLTE on the emitted spectra are smaller than in the atmospheres 
with higher $\Teff$. In Figs.~\ref{low-teff} and \ref{low-teff1} we show the 
UV (panel a) and optical (panel b) spectra of models with $\Teff=10,000\,$K 
(\ref{low-teff}) and $\Teff=15,000\,$K (\ref{low-teff1}). In the top plot of 
each graph, we display the synthetic spectra computed with Fe~II LTE (dotted 
curves) and Fe~II NLTE treatment (full curves). In order to show the 
differences between the spectra in greater detail, we plot in the bottom part 
of each panel the relative differences between the spectra of the LTE and NLTE 
Fe~II models. Note that the relative differences are plotted such that a 
positive value indicates that the flux of the LTE model is larger than the 
flux emitted by the NLTE model and vice versa. 

 We consider first the UV spectra. In general, the Fe~II NLTE spectrum shows 
slightly less flux in this spectral region than the corresponding Fe~II LTE 
spectrum. A number of the Fe~II features and blends, in particular the 
$\lambda 2640$ feature (this is {\em not} an emission line for these effective 
temperatures, cf.\ \cite{novaphys}) and the Fe~II--Mg~II complex at $\lambda 
2800\ang$, are stronger in the Fe~II NLTE spectrum. For smaller wavelengths, 
i.e., the IUE SWP range $1000\ang-2200\ang$, the differences between the two 
spectra can be as much as a factor of 10. However, the total amount of flux 
emitted in the SWP range is small. The differences between LTE-\fep\ and NLTE-
\fep\ models are also relatively small in the IUE LWP range ($2300-3200\ang$). 
but  the fit to observed nova spectra is improved with the NLTE-\fep\ models 
(Schwarz \etal, in preparation). 

The changes between the LTE and NLTE-\fep\ spectra are larger in the optical.
In general, the \fep-lines are {\em stronger} in the NLTE models than in the
LTE models. The relative difference can be as much as 80\% for the \fep-
multiplet lines just longward of H$\beta$. These are significant changes, and
could be interpreted as abundances changes of iron by more than a factor of
two. Therefore, NLTE effects for \fep must be included in the analysis of the
optical spectra of novae even in the earliest stages of its evolution. The
fundamental differences between the effects of NLTE on the \fep\ lines in the
UV and the optical spectral ranges show that a very detailed and careful
treatment of \fep-NLTE is necessary for reliable results. Simplified
treatments, e.g., assuming ionization corrections but ``internal'' LTE (i.e.,
assuming that $b_i = b_j$ for $i \ne j$) for Fe~II, may lead to wrong
conclusions.

\subsubsection{Nova atmospheres with $\Teff \ge 20,000\,$K}

The effect of \fep-NLTE on the synthetic spectra is very different for higher 
effective temperatures. In Figs.~\ref{high-teff} and \ref{high-teff1} we show 
the UV (panel a) and optical (panel b) spectra for models with 
$\Teff=20,000\,$K (Fig.~\ref{high-teff}) and $\Teff=25,000\,$K 
(Fig.~\ref{high-teff}), the different parts of the figures and the meaning of 
the line styles is the same as above. The changes in the UV spectra are now 
very large. The \fep-LTE spectra predict very strong \fep-lines. In fact, they 
would be the strongest lines in the UV spectrum if the assumption of LTE was 
valid for \fep\unskip. That this is not the case is obvious from the 
corresponding NLTE-\fep\ spectrum. The \fep\ lines that dominated the UV 
spectrum in the LTE-\fep\ model are reduced to near invisibility in the NLTE 
model. This is especially true for the \fep\ lines in the wavelength range 
from $2400$ to $2800\ang$ (including the \fep-Mg~II blend at $2800\ang$). It 
is clear that the assumption of LTE for \fep\ has completely broken down and 
that NLTE effects must be considered in order to correctly interpret the UV 
spectrum. The relative differences can be as much as factor of 5 and also the 
changes in the appearance of the spectrum are significant. 

 In contrast, the differences between the optical spectra of LTE-\fep\ and 
NLTE-\fep\ are very small, at most 6\%, for the model with $\Teff=25000\,$K. 
For this effective temperature, the optical \fep\ lines are already weak and 
thus the NLTE effects on \fep\ cannot change the emitted spectrum 
significantly. The model with $\Teff=20,000\,$K shows differences of up to 
20\%. As for the UV spectrum, and contrary to the situation at lower $\Teff$, 
the NLTE \fep\ lines are now weaker than in the LTE case. The NLTE effects on 
the \fep\ line formation are also significant in the optical ranges and, 
again, could be {\em wrongly} interpreted as abundances changes. In the case 
of novae, where abundance changes of iron are possible due both to mixing of 
core material with accreted material and to CNO nuclear reactions (actually 
depletion of hydrogen), this is a very important fact. 

\subsection{Result of test calculations}

 We mentioned earlier in this paper that additional wavelength points for the
\fep\ lines below $3500\ang$ are not required for the model calculations. This
is because the UV lines of \fep\ occur in dense clusters which can be modeled
accurately by our already fine wavelength grid. To verify this, we have
performed full model calculations with 5 additional wavelength points for
every primary \fep line, leading to a total of 90,000 wavelength points. 
We have compared the ground state departure coefficients
for the $15,000$ and the $25,000\,$K models computed with
either 90,000 wavelength points (5 points per primary NLTE line plus default
wavelength points) or 25,000 wavelength points (default
wavelength points and no additional wavelength points for \fep\ lines below
$3500\ang$). 
The differences in the ground state departure
coefficients are less
than the accuracy required for the iterations (1\%), and it is clear that the
departure coefficients are independent of the number of wavelength
points (for our choice of the regular wavelength grid). Finally, in
Fig.~\ref{nosamp} we compare the synthetic spectra computed for the 90,000
wavelength point models (full curves) with the spectra computed for the 25,000
wavelength point models (dotted lines) for the $\Teff=15,000\,$K model. 
The differences are very small in
these spectra and can hardly be seen in the plots. This shows that our smaller
wavelength grid can be used for the model calculations, fully resolving the 
cores and the wings of the lines, resulting in savings
of a factor of 3 in total computer time. The smaller wavelength grid, however,
can only be used because of the properties of the \fep\ ion, for other large
NLTE ions, e.g., Ti~I (Hauschildt \etal, in preparation), the large wavelength
grid may be required for model calculations. This results only in longer model
calculations (a complete model calculation (20 iterations) with 90,000
wavelength points takes about 12,000 seconds of CPU time on one Cray C90
processor) but does {\em not} require more storage or memory. Therefore, even
model calculations with 90,000 wavelength points do not pose any technical
problem and can be done on high-end workstations.

\section{Summary and Conclusions}

 In this paper, we have for the first time reported on nova model atmosphere 
calculations done with a new and detailed NLTE treatment for \fep. Our model 
atom includes 617 levels and 13675 primary transitions. The model atmospheres 
are fully self-consistent solutions of the nongrey radiative transfer and 
energy equations and the NLTE rate equations (including a number of other NLTE 
species in addition to Fe~II). We have compared the results of 
model calculations with 90,000 and 25,000 wavelength points (the latter 
especially selected to guarantee good coverage of the \fep lines) and found 
insignificant differences. 

 We find that the NLTE effects for \fep\ are extremely important in all nova 
models that we have considered. The changes in the synthetic spectra are 
largest for higher effective temperatures ($\Teff \ge 20,000\,$K) and, in 
particular, for the ultraviolet. For smaller effective temperatures, the 
changes in the synthetic UV spectra are smaller, however, they are significant 
for the analysis of optical nova spectra.  NLTE effects are also significant 
for the formation of optical \fep\ lines and they must be included in
interpreting the optical spectra of novae. 

 One of the major effects of NLTE \fep\ line formation is a strong
overionization from \fep\ to Fe~III, caused by the strong UV radiation fields
present in nova atmospheres. These radiation fields keep Fe~III from
recombining to \fep\ throughout the atmosphere. This is an important NLTE
effect and LTE-\fep\ models predict too much \fep\ in the line forming
regions of nova winds.

 In subsequent papers (Schwarz \etal, in preparation) we will show that
inclusion of \fep-NLTE leads to improved fits to the observed optical and UV
spectra of novae. 

\acknowledgments

It is a pleasure to thank H. St\"orzer, J. Krautter, and S.
Shore for stimulating discussions.
This work was supported
in part by NASA LTSA grants to Arizona State University, by NASA grant
NAGW-2999; as well as NSF grants to Wichita State University.
Some of the calculations presented in this paper were
performed at the San Diego Supercomputer Center (SDSC), supported by the
NSF, and at the NERSC, supported by the U.S. DoE, we thank them for a
generous allocation of computer time.
The development of \phoenix's opacity database was sponsored in parts
by a grant to FA from the American Astronomical Society.

\bibliography{yeti,opacity,novae,ltsa}

\clearpage
\section{Figure Captions}
\begin{figure}
\caption[]{\label{linefig}The number of overlapping lines in the co-moving
frame for a nova atmosphere model with $\Teff=20,000\,$K. We show the number
of lines that lie within a $\pm 2$ Doppler-width interval at each wavelength
point. In panel (a) only lines that are stronger than the local b-f continuum
are included, whereas in panel (b) all lines that are stronger than $10^{-4}$
times the local b-f continuum are included. We have used a micro-turbulent
velocity of $\xi=2\kms$ in this calculation.}
\end{figure}


\begin{figure}
\caption[]{\label{bground}The run of the ground state departure
coefficients $b_1$ for the NLTE species as functions of $\tstd$ (the b-f
optical depth at $5000\ang$). The different panels show the results obtained
for models with different effective temperatures, as indicated. The common
model parameters are $L=50,000\Lsun$, $N=3$, $\vmax=2000\kms$, and solar
abundances.}
\end{figure}

\begin{figure}
\caption[]{\label{fe-ion}Ionization structure of iron as function of
$\tstd$ (the b-f optical depth at $5000\ang$). Panel (a) shows the results
obtained with Fe~II treated in NLTE, whereas panel (b) shows the results for
LTE Fe~II. The NLTE overionization keeps the Fe~III from recombining to Fe~II
in the outer layers of the nova atmosphere. Note the large number
of ``Str\"omgren Spheres'' of various ionization stages of iron in the inner
parts of the nova atmosphere.}
\end{figure}

\begin{figure}
\caption[]{\label{ion-struc}Ionization structure of the $\Teff=20,000\,$K nova
atmosphere model as function of
$\tstd$ (the b-f optical depth at $5000\ang$). The graph shows the ratio of
the partial pressures to the gas pressure for the 12 most abundant species.}
\end{figure}

\begin{figure}
\caption[]{\label{nefrac}The partial pressure ratio $p({\rm Fe~II})/p({\rm
Fe~III})$ (curve marked with plus-symbols, left hand y-axis) and the ratio
$p_{\rm elec}/\pgas$ (dotted curve, right hand y-axis) as functions of
standard optical depth $\tstd$. The
sudden changes in the electron pressure are caused mainly by the two helium
ionization zones in combination with the rapid changes in the iron ionization
degree,  which is caused by the behavior of the departure coefficients.}
\end{figure}

\begin{figure}
\caption[]{\label{allbi} The departure coefficients for all NLTE level of \fep\
as functions of the standard optical depth $\tstd$ for all 4 models. The
plots demonstrate that the NLTE effects become more important with higher
effective temperatures. In addition, the
departure coefficients are closer to unity for the higher lying levels.}
\end{figure}

\begin{figure}
\caption[]{\label{uv1}The ratio of the line source function to the
local Planck function for the UV1 multiplet ($\rm a ^6D$ to $\rm z
^6D^o$) as functions of the
standard optical depth. The two panels are for models with
$\Teff=15000\,$K (a) and $\Teff=25000\,$K (b).}
\end{figure}

\begin{figure}
\caption[]{\label{42}The ratio of the line source function to the
local Planck function for multiplet 42 ($\rm a ^6S$ to $\rm z
^6P^o$) as functions of the
standard optical depth. The two panels are for models with
$\Teff=15000\,$K (a) and $\Teff=25000\,$K (b).}
\end{figure}

\begin{figure}
\caption[]{\label{from}The ratio of the line source function to the
local Planck function for lines originating from
the ground term $\rm a ^6D$ (panel a), the metastable $\rm a ^4F$ term
(panel b), and the higher $\rm z ^4P^o$ term (panel c) for both the
$15,000$ and the $25,000\,$K models.}
\end{figure}

\begin{figure}
\caption[]{\label{low-teff} The UV and
optical spectra of models with $\Teff=10,000\,$K.
We show the
synthetic spectra for the full Fe~II NLTE treatment (full curves) and for LTE
Fe~II (dotted curves). The bottom parts of each panel we show the ratio of the
LTE-Fe~II and NLTE-Fe~II spectra to emphasize the difference between the
spectra.}
\end{figure}

\begin{figure}
\caption[]{\label{low-teff1} The UV and
optical spectra of a model with $\Teff=15,000\,$K.
We show the
synthetic spectra for the full Fe~II NLTE treatment (full curves) and for LTE
Fe~II (dotted curves). The bottom parts of each panel we show the ratio of the
LTE-Fe~II and NLTE-Fe~II spectra to emphasize the difference between the
spectra.}
\end{figure}

\begin{figure}
\caption[]{\label{high-teff} The UV and
optical spectra of models with $\Teff=20,000\,$K.
In the top parts of the graphs we show the
synthetic spectra for the full Fe~II NLTE treatment (full curves) and for LTE
Fe~II (dotted curves). The bottom parts of each panel we show the ratio of the
LTE-Fe~II and NLTE-Fe~II spectra to emphasize the difference between the
spectra.}
\end{figure}

\begin{figure}
\caption[]{\label{high-teff1} The UV and
optical spectra of models with $\Teff=25,000\,$K.
In the top parts of the graphs we show the
synthetic spectra for the full Fe~II NLTE treatment (full curves) and for LTE
Fe~II (dotted curves). The bottom parts of each panel we show the ratio of the
LTE-Fe~II and NLTE-Fe~II spectra to emphasize the difference between the
spectra.}
\end{figure}

\begin{figure}
\caption[]{\label{nosamp}The synthetic spectra computed for a model with
$\Teff=15000\,$K with 90,000 wavelength points (full curves)
and 25,000 wavelength points (dotted curves) in the ultraviolet (top plots)
and the optical (bottom plots).
The plots demonstrate that the results of the model calculations do
not significantly depend on the number of wavelength points used in
the models as long as enough wavelength points are used 
(in this case $\ga 25,000$ points)
and are placed correctly so that the wavelength space is saturated.
}
\end{figure}

\end{document}